\documentclass[prb,amsmath,superscriptaddress,twocolumn]{revtex4}
\usepackage{bm}
\usepackage{amssymb}
\usepackage{graphicx}

\newcommand{\up}{\uparrow}
\newcommand{\down}{\downarrow}

\def\nab{{\mbox{\boldmath{$\nabla$}}}}
\def\r{{\mbox{\boldmath{$\rho$}}}}

\def\sig{{\mbox{\boldmath{$\sigma$}}}}

\begin{document}

\title{Normal persistent currents in  proximity-effect  bilayers}

\author{O. Entin-Wholman}
\email{oraentin@bgu.ac.il}

\altaffiliation{Also at Tel Aviv University.}
\affiliation{Department of Physics and the Ilse Katz Center for
Meso- and Nano-Scale Science and Technology, Ben Gurion
University, Beer Sheva 84105, Israel}

\author{H. Bary-Soroker}

\affiliation{Department of Physics, Ben Gurion University, Beer
Sheva 84105, Israel}

\author{A. Aharony}

\altaffiliation{Also at Tel Aviv University.}

\affiliation{Department of Physics and the Ilse Katz Center for
Meso- and Nano-Scale Science and Technology, Ben Gurion
University, Beer Sheva 84105, Israel}

\author{Y. Imry}

\affiliation{Department of Condensed Matter Physics,  Weizmann
Institute of Science, Rehovot 76100, Israel}

\author{J. G. E. Harris}

\affiliation{Department of Physics, Yale University, New Haven, CT 06520, USA}

\affiliation{Department of Applied Physics, Yale University, New Haven, CT 06520, USA}

\date{\today}


\begin{abstract}
We calculate the contribution of superconducting fluctuations to
the mesoscopic persistent current of an ensemble of  rings, each
made of a superconducting layer in contact with a normal one, in
the Cooper limit. The superconducting transition temperature of
the bilayer decays very quickly with the increase of the relative
width of the normal layer. In contrast, when the Thouless energy
is larger than the temperature then the suppression of the
persistent current with the increase of this relative width is
much slower than that of the transition temperature. This effect
is similar to that predicted for magnetic impurities, although the
proximity effect considered here results in pair-weakening as
opposed to pair-breaking.
\end{abstract}

\maketitle

\section{Introduction}

The average persistent  current    \cite{BUTTIKER,BOOK} of a large
number of mesoscopic metallic rings can be used to deduce the sign
and  the magnitude of electron-electron interactions in the metal
forming the rings. The size of the average current is expected to
increase with the strength of the interactions, and its sign
reflects the nature of the interactions: the magnetic response at
low flux is paramagnetic (diamagnetic) when the electronic
interactions are repulsive  (attractive).\cite{AEEPL,AEPRL}  For a
large ensemble of rings, the current is expected to be periodic in
the magnetic flux, with the period corresponding to one half of
the flux quantum, $h/2e$. The theoretical analysis of Refs.
\onlinecite{AEEPL} and \onlinecite{AEPRL} was motivated in part by
early measurements of the average persistent current in an
array of $10^6$ copper rings, \cite{LDDB} whose sign and magnitude
could not be accounted for by  noninteracting electrons alone and
therefore should be affected by electronic interactions. These
experiments confirmed the above periodicity, also suggesting that
the average magnetic response is induced by interactions.
\cite{AEEPL,AEPRL}
 Similar results were later observed on an array of $10^5$ GaAs rings \cite{reulet} and on
 an array of $10^5$ silver rings.
\cite{DBRBM}  In contrast,
measurements on  single
rings\cite{webb91,10,11,Moler}
 showed the $h/e$
periodicity. In an array of 30 gold rings\cite{JMKW} both the
$h/2e$ and the $h/e$ harmonics were observed. In this paper the
authors were unable to say whether the $h/2e$ signal was the
second harmonic of the typical contribution or the  first harmonic
of an average contribution. Overall, the sign of the $h/2e$
harmonic measured on metallic rings seems to indicate that the
low-flux response is diamagnetic, \cite{DBRBM,JMKW}  implying
attractive interactions. Recently, Bleszynski-Jayich {\it  et al.}\cite{JGEH}
 found that the average current in
aluminum rings, subject to high magnetic fields, is negligible,
but typical mesoscopic fluctuations remain almost unaffected. 
It seems that these latter experiments can be explained within
the framework of noninteracting electrons. \cite{EG,HAMUTAL3}

Interestingly enough, it turned out that the {\it bona fide}
values of the attractive interactions required to explain the
persistent-current data of the copper \cite{LDDB}  ensemble for
example, would have implied that this metal is superconducting at
measurable temperatures, of the order of 1mK. In fact, early experiments on the
magnetic response \cite{GD0} and on the thermal
conductivity\cite{GD2} of proximity-effect systems, whose normal
parts were copper and silver, also indicated a minute attractive
interaction in these metals.\cite{GD1}  However, these early measurements allowed a
broad range for the magnitude of this interaction, and therefore did not open a discussion of
the reasons for the absence of superconductivity  in experiments on these metals.
The latter puzzle became obvious only after the measurements of the persistent current on
copper, which requires a transition temperature of 1mk.
Superconductivity
has not been detected also in gold and silver, and this fact has remained unexplained for many years. A
possible explanation for this apparent puzzle \cite{HELENE} was
offered in Refs. ~\onlinecite{HAMUTAL1} and
~\onlinecite{HAMUTAL2}, which argued (for the first time) that the existence of
(seemingly unavoidable\cite{BIRGE}) tiny amounts of magnetic
impurities may detrimentally affect superconductivity in such
metals, reducing their transition temperatures to undetectable,
even zero,  values, while leaving the persistent current almost
unharmed. This stems from the disparity of the energy scales
determining the renormalized electronic interaction pertaining  to
each phenomenon. The interaction-induced persistent current is
proportional to the renormalized interaction on the scale of the
Thouless energy, $E^{}_{c}=\hbar D/L^{2}$ (where $D$ is the
diffusion coefficient and $L$ is the circumference of the ring).
Superconductivity is lost, however, when the spin-flip rate  of
the magnetic impurities, $\hbar/\tau^{}_{s}$ (in units of energy),
becomes comparable to the {\em bare} transition temperature of the
material (in the absence of any pair-breaking or pair-weakening
agents), $T^S_{c0}$. In other words, the actual superconducting
transition temperature $T^S_{c}$ is determined by the {\it
renormalized} interaction, on the scale of $\max[T^S_{c0},\hbar/\tau^{}_{s}]$. It
follows that a concentration of magnetic impurities  such that
\begin{align}
 k^{}_B T^{S}_{c0}\lesssim \hbar/\tau^{}_{s}\lesssim E^{}_{c}\
,\label{ineq}
\end{align}
will hardly affect the magnitude of the persistent current,
concomitantly suppressing the superconducting transition
temperature (below we often use units in which $\hbar=k_{\rm
B}=1$). Indeed, detailed analysis \cite{HAMUTAL2} of the
persistent current data reported in Refs. \onlinecite{LDDB} and
\onlinecite{DBRBM} led to the conclusion that $T^S_{c0}$ of copper
(gold) is in the mK (a fraction of mK) range.

\begin{figure}[ hbtp]
\includegraphics[width=5cm]{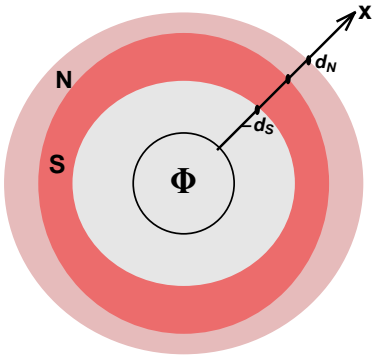}
\caption{(color online) Proximity-effect ring in which the width of the strong (weak)
 superconductor is $d_{S}$ ($d_{N}$), threaded by a magnetic flux $\Phi$ (measured in units of the flux
 quantum).
 }\label{first}
\end{figure}

This theoretical picture can be tested, for instance, by
investigating rings made of known
low-superconducting-transition-temperature materials, in which a
controlled concentration of pair breakers have been
added.\cite{JGEH} It has also been noted that the magnetic flux
itself acts as a pair breaker, causing a periodic decrease of the
transition temperature but a lesser decrease in the persistent
current. \cite{oreg} It is interesting to check whether there
exist other situations where the superconducting transition
temperature is lowered by some pair-breaking or pair-weakening
mechanism, but the (superconducting fluctuation-induced)
persistent current remains large far above this transition
temperature. In the present paper we consider this question for
superconducting-normal ($SN$) bilayers, e.g. made of Al and
Cu.\cite{HELENE}  Bilayers made of Al and Ag might even be better,
as they avoid magnetic impurities. The `normal' metal could also
be a weaker superconductor, with a lower transition temperature.
The proximity effect is known to cause a decrease of the
transition temperature of the bilayer with the relative thickness
of the $S$ layer,\cite{DEGENNES,ANI} and it is interesting to find
out what happens to the persistent current, which is induced by
superconducting fluctuations.
 This possibility is in particular intriguing: unlike the
 magnetic impurities,
 the proximity effect  is not a {\it bona fide}
pair-breaker, since time-reversal invariance is not broken by it.
The proximity effect just leads
 to pair-weakening, by `diluting' the superconducting fraction. \cite{FULDE}

Here we present a calculation of the disorder-averaged persistent
current averaged over an ensemble of bilayer rings, each  having
the geometry depicted in  Fig. \ref{first}. These rings consist of
two adjacent metallic rings, with different transition
temperatures. The area inside the rings is penetrated by a
magnetic flux $\Phi$, which is measured in units of the flux
quantum $hc/e$. Below we use the subscript $S$ for quantities
characterizing the layer with the higher transition temperature,
and the subscript $N$ for the quantities belonging to the  other
one, which may or may not be a superconductor. For simplicity, we
confine our calculation to bilayers in the Cooper limit:
\cite{DEGENNES}  this limit is reached when the width of each of
the layers, $d_{S}$ or $d_{N}$, is much smaller than the
respective    coherence length. \cite{COM2}  Our aim is to explore
the possibility to deduce the scale of the renormalized electronic
interaction by analyzing concomitantly the superconducting
transition temperature and the (superconducting)
fluctuation-induced average persistent current. In other words, we
examine the persistent current as a function of the $N-$slab
relative thickness, and find parameter regimes where it is
affected much less than the transition temperature.

Since pair breakers, notably magnetic impurities, seem to be
ubiquitous in several of the metals used in the persistent-current
measurements, it  is interesting to investigate their effect in a
proximity-effect configuration. For instance, it is plausible that
in Al/Cu rings, the copper (the $N-$slab in our notations) may
well include a tiny amount of magnetic impurities. We therefore
include scattering off such impurities in our expressions.

\begin{figure}[ hbtp]
\includegraphics[width=7cm]{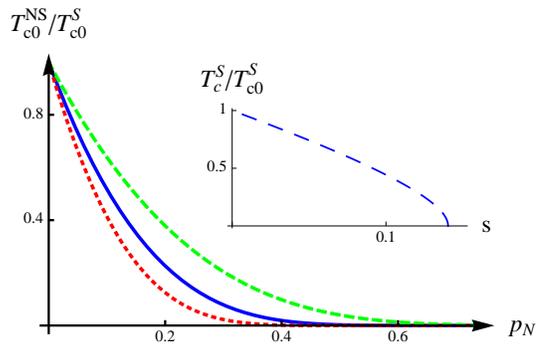}
\caption{(color online) The transition temperature of a
proximity-effect sandwich in the Cooper limit [Eq. (\ref{TNS01})]
as a function of the weighed width of the normal layer for
$\lambda_{N}=0,~0.05$, and $-0.05$ (solid, dashed, and dotted
curve, respectively). In all curves $\lambda_{S}=0.17$. The inset
depicts the Abrikosov-Gorkov expression, Eq. (\ref{TAG}), for the
reduction of the transition temperature by pair breakers.
 }\label{second}
\end{figure}

The transition temperature of an $SN$ proximity bilayer in the the
Cooper limit is known\cite{DEGENNES,ANI} to be determined by the
effective (dimensionless) electronic coupling, $\lambda^{}_{NS}$, which is the
weighted sum of the effective couplings of the separate slabs,
$\lambda_{S}$ (which is positive, since the $S-$slab is
superconducting) and $\lambda_{N}$ (which may take both signs):
\begin{align}
\lambda ^{}_{NS}=p^{}_{N}\lambda^{}_{N}+p^{}_{S}\lambda^{}_{S}\
,\label{COUP}
\end{align}
with
\begin{align}
p^{}_{N(S)}=d^{}_{N(S)}\widetilde{\cal
N}^{}_{N(S)}/\widetilde{\cal N}^{}_{\rm eff}\ ,\label{PNS}
\end{align}
where $\widetilde{\cal N}_{N(S)}$ denotes the density of states at
the Fermi energy {\em per unit length}  of the normal
(superconducting) layer, and
\begin{align}
\widetilde{\cal N}^{}_{\rm eff}=\widetilde{\cal
N}_{N}^{}d^{}_{N}+\widetilde{\cal N}^{}_{S}d^{}_{S}\ .\label{NEFF}
\end{align}
The mean-field transition temperature,   $T_{c0}^{NS}$,  of the
bilayer (without magnetic impurities) is then given by
\cite{DEGENNES,ANI}
\begin{align}
\frac{1}{\lambda^{}_{NS}}=\Psi \Bigl (\frac{1}{2}
+\frac{\omega^{}_{D}}{2\pi T_{c0}^{NS}}\Bigr )- \Psi \Bigl
(\frac{1}{2}\Bigr )\ ,\label{TC0NS}
\end{align}
where $\Psi$ is the digamma function whose asymptotic expansion,
valid for large arguments, is given by
\begin{align}
\lim_{z\to\infty}\Psi (z)\to\ln z\ .\label{ASYM}
\end{align}
 The
Debye frequency $\omega^{}_{D}$ in Eq. (\ref{TC0NS}) (assumed to be
identical for both slabs) marks the upper cutoff on the effective
interactions. The result (\ref{TC0NS}) is obtained assuming that
the two layers are in a good electrical contact. (The effect of a
barrier between the two layers has been considered by McMillan.
\cite{MAC}) In the limit $\omega^{}_{D}\gg T^{NS}_{c0}$ one may
use Eq. (\ref{ASYM}) to obtain
\begin{align}
\ln\frac{T^{NS}_{c0}}{T^{S}_{c0}}=-\Bigl
(1-\frac{\lambda^{}_{N}}{\lambda^{}_{S}}\Bigr )
\frac{p^{}_{N}}{\lambda^{}_{N}p_{N}^{}+\lambda^{}_{S}p^{}_{S}}\
,\label{TNS01}
\end{align}
where $T^{S}_{c0}$  is the bulk transition temperature of the
clean $S-$slab. When the $N-$ slab is also superconducting (i.e.
$\lambda^{}_N>0$), $T^{NS}_{c0}$ remains finite for all $p^{}_N$
(although quite small for large $p^{}_N$ and small
$\lambda^{}_N$). However, when $\lambda^{}_N\leq 0$, the
transition temperature of the bilayer $T^{NS}_{c0}$ approaches
zero at a {\it quantum critical point},
$p^{}_N=1/(1-\lambda^{}_N/\lambda^{}_S)$. The approach is
exponential, with {\it zero} slope (see Fig. \ref{second}). In
practice, $T^{NS}_{c0}$ becomes very small for $p^{}_N\gtrsim
1/2$. In some sense, this inequality  replaces the left hand side
of Eq. (\ref{ineq}). As we show below, the persistent current
remains rather large even in this regime.

The effect of  genuine pair-breaking mechanisms on the transition
temperature  was considered a long time ago. The seminal paper of
Abrikosov and Gorkov \cite{AG} found that  the transition
temperature $T^{S}_{c0}$ is reduced by magnetic impurities to
$T^S_c$,
\begin{align}
\ln\frac{T^{S}_c}{T^{S}_{c0}}=\Psi\Bigl(\frac{1}{2}\Bigr)-\Psi\Bigl(\frac{1}{2}+\frac{s
T^S_{c0}}{ T^S_c}\Bigr)\ ,\label{TAG}
\end{align}
where $s=1/(2\pi T^S_{c0}\tau^{}_s)$. This expression is shown in
the inset in Fig. \ref{second}. Unlike $T^{NS}_{c0}$, $T^S_c$
approaches zero at $s=\exp[\Psi(1/2)]=1/(4\gamma^{}_E)\approx
0.140365$ (namely at $\hbar/\tau^{}_s \sim 0.9 T^S_{c0}$), with a
{\it finite} slope. Here, $\gamma^{}_E$ is the Euler constant.
This difference in slope between the two mechanisms probably
reflects the difference between pair-weakening and pair-breaking.
\cite{FULDE}

 In a complete analogy with Eq. (\ref{TAG}), a small amount of pair-breaking impurities in the $N-$slab
lowers the transition temperature of the sandwich from
$T_{c0}^{NS}$ to $T_{c}^{NS}$, given by \cite{ANI}
\begin{align}
\frac{1}{\lambda^{}_{NS}}&=\Psi \Bigl
(\frac{1}{2}+\frac{\omega^{}_{D}}{2\pi T_{c}^{NS}}\Bigr ) - \Psi
\Bigl (\frac{1}{2}+\frac{p^{}_{N}}{2\pi
T^{NS}_{c}\tau_{s}^{}}\Bigr )\ ,\label{TCNS}
\end{align}
generalizing Eq. (\ref{TC0NS}).

The rest of this paper describes the calculation of the average
persistent current, pertaining to a large ensemble of bilayers.
Section \ref{PSSA} outlines the derivation of the effective
Ginzburg-Landau theory for this case, with some technical details
given in Appendix \ref{DETC}. Some quantitative results are
presented in Sec. \ref{RES}. Since the fluctuations are calculated
within the high-temperature Gaussian approximation, which is valid
only above the Ginzburg critical regime, Sec. \ref{Gin} presents a
critical discussion of this regime. That section also contains our
conclusions.

\section{The persistent current of a proximity-effect sandwich}

\label{PSSA}

Here we present a microscopic derivation of the free energy which
determines the superconducting fluctuations. The Hamiltonian of
the bilayer is similar to that used in Refs. \onlinecite{HAMUTAL1}
and \onlinecite{HAMUTAL2},
\begin{align}
{\cal H}=\int d{\bf r}{\cal H}({\bf r})\ ,
\end{align}
with
\begin{align}
{\cal H}({\bf r})&=\sum_{\sigma \sigma
'}\psi_{\sigma}^{\dagger}({\bf r})
{\cal H}^{(0)}_{\sigma\sigma '}({\bf r})\psi^{}_{\sigma '}({\bf r})\nonumber\\
&-V({\bf r})\psi^{\dagger}_{\up}({\bf
r})\psi^{\dagger}_{\down}({\bf r})\psi^{}_{\down}({\bf r})
\psi^{}_{\up}({\bf r})\ , \label{H}
\end{align}
where $\psi_{\sigma}^{\dagger}({\bf r})$ creates an electron with
spin $\sigma$ at ${\bf r}$. The interaction $V({\bf r})$ depends
on the spatial coordinate $x$ (see Fig. \ref{first}),
\begin{align}
V(x)&=\lambda^{}_{S}/{\cal N}^{}_{S}\ ,\ \ \ -d^{}_{S}\leq x\leq 0\nonumber\\
V(x)&=\lambda^{}_{N}/{\cal N}^{}_{N}\ ,\ \ \  0\leq x\leq
d^{}_{N}\ .
\end{align}
(Note that here the ${\cal N}$'s are the densities of states {\em
per unit volume} of the two layers.) The single-particle part of
the Hamiltonian (\ref{H}) reads
\begin{align}
{\cal H}^{(0)}_{\sigma\sigma '}({\bf r})=\delta^{}_{\sigma ,\sigma
'}{\cal H}^{}_{0}({\bf r}) +u^{}_{\sigma\sigma '}({\bf r})\
,\label{H0}
\end{align}
where
\begin{align}
{\cal H}^{}_{0}= [-i\nab+(e/c){\bf A}({\bf r})]^{2}/(2m)-\mu\
,\label{Hzero}
\end{align}
 and $\mu$ is the chemical potential.
 With the choice ${\bf A}={\bf B}\times{\bf r}/2$, the vector potential ${\bf A}$ points along the
 circumference of the ring in the anticlockwise direction.
The disorder potential is $u({\bf r})\equiv
u^{}_{1}+u^{}_{2}\sig\cdot{\bf S}$, yielding scattering off
nonmagnetic impurities (scaled by $u^{}_{1}$) as well as off
magnetic impurities (scaled by $u^{}_{2}$, ${\bf S}$ denotes the
magnetic impurity spins. The impurities are modeled by point-like
scatterers \cite{AGD}).

The quantum partition function ${\cal Z}$ is\cite{ALTLAND}
\begin{align}
{\cal Z}=\int {\cal D}\{\psi ({\bf r},\tau ),\overline{\psi}({\bf
r},\tau )\}\exp [-{\cal S}]\ ,\label{ZZ}
\end{align}
where the action ${\cal S}$ is
\begin{align}
{\cal S}=\int d{\bf r}\int_{0}^{\beta}d\tau \Bigl
(\sum_{\sigma}\overline{\psi}^{}_{\sigma} ({\bf r},\tau )
\frac{\partial}{\partial\tau} \psi^{}_{\sigma} ({\bf r},\tau
)+{\cal H}({\bf r},\tau )\Bigr )\ ,\label{S}
\end{align}
and $\beta=1/T$. Here, the annihilation and creation  field
operators  in the Hamiltonian  (\ref{H}) ($\psi$ and $\psi^\dagger$) are replaced by the
spinor Grassmann variables $\psi ({\bf r},\tau )$ and
$\overline{\psi}({\bf r},\tau )$, respectively.

Since the calculation of the partition function is rather
technical, we present it in Appendix \ref{DETC}. We first perform
this analysis in the absence of the magnetic flux. A
Hubbard-Stratonovich transformation replaces the Grassmann
variables $\psi$ and $\overline{\psi}$ by complex bosonic
variables $\Delta({\bf r},\tau)$ and $\Delta^\ast({\bf r},\tau)$
(which are now functions of the imaginary time $\tau$), and the
action is expanded in powers of these variables. The Gaussian
approximation uses only the quadratic terms in this expansion. In
the Cooper limit, the Fourier transformed bosonic variables take
only two values as a function of $x$, namely $\Delta^{}_S({\bf
q},\nu)$ for $-d^{}_S<x<0$ and $\Delta^{}_N({\bf q},\nu)$ for
$0<x<d^{}_N$, where ${\bf q}$ is a two-dimensional vector
perpendicular to ${\hat x}$. The quadratic action then becomes
\begin{align}
{\cal S}^{}_2&=\beta\widetilde{\cal N}^{}_{\rm eff}\sum_{\bf
q}\sum_{\nu} \Bigl (a^{}_N|\Delta^{}_{N}({\bf q},\nu )|^{2}
+a^{}_S|\Delta^{}_{S}({\bf q},\nu )|^{2}\nonumber\\
&\hspace{1cm}-c\bigl[\Delta^{\ast}_{N}({\bf
q},\nu)\Delta^{}_{S}({\bf q},\nu )+c.c.\bigr]\Bigr )\
,\label{ACT2}
\end{align}
where $\widetilde{\cal N}_{\rm eff}$ is given in Eq. (\ref{NEFF}),
and
\begin{align}
a^{}_{N(S)}=\frac{p^{}_{N(S)}}{\lambda^{}_{N(S)}}-p^{2}_{N(S)}\gamma\
,\ \ \  c=p^{}_{S}p^{}_{N}\gamma\ .\label{ANS}
\end{align}
Here, $p^{}_{N(S)}$  are  given in Eqs. (\ref{PNS}).  The function
$\gamma ({\bf q},\nu, T )$ is given by
\begin{align}
\gamma ({\bf q},\nu ,T)=\pi T\sum_{\omega}\bar{\gamma}({\bf q},\nu
,\omega)\ , \label{P}\end{align}
where
\begin{align}
\bar{\gamma}({\bf q},\nu ,\omega)=\Bigl (|\omega| +\frac{|\nu
|}{2} +\frac{p^{}_{N}}{\tau^{}_{s}} +\frac{1}{2}D^{}_{\rm eff}{\bf
q}^{2}\Bigr )^{-1}\ ,\label{GAM}
\end{align}
$\omega=\pi T(2m+1)$ and $\nu=2\pi T\ell$ (with integer $m$ and
$\ell$) are the fermionic and bosonic Matsubara frequencies and
$D^{}_{\rm eff}$ is the effective diffusion coefficient of the
double layer,
\begin{align}
D^{}_{\rm eff}=p^{}_{N}D^{}_{N}+p^{}_{S}D^{}_{S}\ .\label{DEFF}
\end{align}
Since the sum over $\omega$ in Eq. (\ref{P}) is cut-off by the
Debye frequency $\omega^{}_{D}$ one finds
\begin{align}
\gamma({\bf q},\nu ,T)=-\Psi\Bigl(\frac{1}{2} +\frac{|\nu|
+D^{}_{\rm eff}{\bf q}^{2}+2p_{N}/\tau_{s}}{4\pi T}\Bigr
)\nonumber\\
+\Psi\Bigl(\frac{1}{2} +\frac{|\nu| +D^{}_{\rm eff}{\bf
q}^{2}+2p_{N}/\tau_{s}+2\omega^{}_D}{4\pi T}\Bigr )\ .\label{gamT}
\end{align}

The bilinear form in Eq. (\ref{ACT2}) is diagonalized by the
transformation (for convenience, we omit  the explicit notations
of ${\bf q}$, $\nu$, and $T$ in part of the expressions below)
\begin{align}
\Delta^{}_{N}=u^{}_{-}\Delta^{}_{-}+u^{}_{+}\Delta^{}_{+}\ ,\ \ \
\Delta^{}_{S}=u^{}_{+}\Delta^{}_{-}-u^{}_{-}\Delta^{}_{+}\
,\label{DELPM}
\end{align}
where
\begin{align}
u^{}_{\pm}=\Bigl
(\frac{1}{2}\pm\frac{a^{}_{N}-a^{}_{S}}{4\kappa}\Bigr )^{1/2}\ ,\
\  \kappa =\sqrt{\Bigl (\frac{a^{}_{N}-a^{}_{S}}{2}\Bigr
)^{2}+c^{2}}\ .\label{UPM}
\end{align}
One then finds
\begin{align}
{\cal S}^{}_2=&\beta\widetilde{\cal N}^{}_{\rm eff}\sum_{\bf
q}\sum_{\nu}\Bigl (a^{}_{-}({\bf q},\nu , T)|\Delta^{}_{-}({\bf
q},\nu )|^2
\nonumber\\
&\hspace{1cm}+a^{}_{+}({\bf q},\nu , T)|\Delta^{}_{+}({\bf q},\nu
)|^2\Bigr )\ ,\label{ACT3}
\end{align}
with \begin{align} a^{}_\pm=(a^{}_N+a^{}_S)/2\pm\kappa\ .
\end{align}

Within this Ginzburg-Landau-like model, the phase transition
occurs when the first coefficient $a^{}_\pm({\bf q},\nu,T)$
vanishes as the temperature $T$ is lowered.  Since
$a^{}_+-a^{}_-=2\kappa>0$, this transition happens when
$a^{}_-(0,0,T)=0$ (while $a^{}_+$ remains positive). Equations
(\ref{ANS}) and (\ref{P}) imply that
\begin{align}
a^{}_+a^{}_-=a^{}_N a^{}_S-c^2= \frac{p^{}_N
p^{}_S\lambda^{}_{NS}}{\lambda^{}_N\lambda^{}_S}\bigl(\lambda^{-1}_{NS}-\gamma({\bf
q},\nu,T)\bigr )\ . \label{a+a-}\end{align} Therefore, at zero
flux the transition occurs at $T^{NS}_c$ which obeys the equation
$\lambda^{-1}_{NS}=\gamma(0,0,T^{NS}_c)$. Using Eq. (\ref{gamT}),
this reproduces Eq. (\ref{TCNS}).

Finally, we incorporate the magnetic flux into the expressions for
the action and for the partition function. To lowest order
(neglecting the effect of the field on the order parameter) it
suffices to replace ${\bf q}$ by\cite{carolimaki}
\begin{align}
{\bf q}\rightarrow {\bf q}+(2e/c){\bf A}\ .
\end{align}
This follows directly from Eq. (\ref{Hzero}), remembering that the
momentum ${\bf q}$ relates to a bosonic Cooper pair. For the
circular geometry at hand, the component of ${\bf q}$ along the
ring circumference, $q^{}_\parallel$, becomes
\begin{align}
q^{}_\parallel=\frac{2\pi}{L}(n+2\Phi)\ ,\label{qpar}
\end{align}
with integer $n$. The transition is then shifted, with\cite{oreg}
\begin{align}
\frac{1}{\lambda^{}_{NS}}=\gamma\bigl(q^{}_\parallel=4\pi\Phi/L,0,T^{NS}_c(\Phi)\bigr)\
, \label{shift}
\end{align} and the persistent current is given by
\begin{align}
I=\frac{e}{2\pi c}\frac{\partial T\ln{\cal Z}}{\partial \Phi} \ .
\end{align}

 Within this Gaussian approximation, the
fluctuations contribution to the partition function can be
obtained straightforwardly. One finds
\begin{align}
{\cal Z}^{}_{\rm fl}&=\prod_{\bf q}\prod_{\nu}
\frac{1}{a^{}_{+}({\bf q},\nu ,T)a^{}_{-}({\bf q},\nu ,T)}\nonumber\\
&\sim\prod_{\bf q}\prod_{\nu}
\frac{1}{\lambda^{-1}_{NS}-\gamma ({\bf q},\nu, T)}\ ,
\label{ZSOF}
\end{align}
where (flux- and temperature-independent) multiplicative factors
have been omitted. Interestingly, this expression for the
partition function has exactly the same form as that found for the
`superconducting' ring in Ref. \onlinecite{HAMUTAL2}. The only
modification is that now $\lambda^{-1}_S$ is replaced by
$\lambda^{-1}_{NS}$. The following calculations thus use the same
calculational techniques employed in that reference.

\section{Results}

\label{RES}

Since the important contribution to the persistent current comes
from the zero transverse mode (perpendicular to the
$x-$direction), \cite{HAMUTAL3,HAMUTAL1,HAMUTAL2} we replace the
sum over ${\bf q}$  by a {\em one-dimensional } summation over the
discrete values of $q^{}_\parallel$, Eq. (\ref{qpar}). Assuming
that the Debye frequency $\omega^{}_D$ is the largest energy in
the problem, the denominator in Eq. (\ref{ZSOF})
becomes\cite{HAMUTAL2}
\begin{align}\label{eq.log}
\lambda^{-1}_{NS}&-\gamma({\bf q},\nu,T)\approx
\ln\Bigl[\frac{T}{T^{NS}_{c0}}\Bigr]
+\Psi[\widetilde{F}(n,\ell)]-\Psi\Bigr[\frac{1}{2}\Bigr]\ ,
\end{align}
\begin{align}
\widetilde{F}(n,\ell)=\frac{1+|\ell|}{2}+\frac{\pi
E^{}_c}{T}(n+2\Phi)^2+\frac{p^{}_N}{2\pi T \tau^{}_s}\ ,
\end{align}
and therefore the persistent current is\cite{HAMUTAL2}
\begin{align}
I=-2eE^{}_c\sum_{n,\ell}\frac{(n+2\Phi)\Psi'(\widetilde{F})}{\ln(T/T^{NS}_{c0})+\Psi(\widetilde{F})-\Psi(1/2)}\
.\label{II1}
\end{align}
(We remind the reader that $E^{}_c=D^{}_{\rm eff}/L^2$ is the
Thouless energy). As shown in Ref. \onlinecite{HAMUTAL2}, this
expression for the persistent current can also be written as a
Poisson summation,
\begin{align}
I=&-4eT\sum_{m=1}^\infty\sin(4\pi m\Phi)\nonumber\\
&\times\sum_\ell\sum_{j=1}^{\infty}\bigl[\exp(2\pi i x^{\ell j}_{\rm
zero})-\exp(2\pi i x^{\ell j}_{\rm pole})\bigr] \ ,\label{III}
\end{align}
where
\begin{align}
x^{\ell j}_{\rm pole/zero}=i m \sqrt{\frac{T}{2\pi
E^{}_c}}\Bigl[1+|\ell|+\frac{p^{}_N}{\pi T \tau^{}_s}-2F^j_{\rm
pole/zero}\Bigr]^{1/2}\ ,\label{xj}
\end{align}
with $F^j_{\rm pole}=-j$ and with $F^j_{\rm zero}$ being the
solution of $\Psi(F^j_{\rm zero})=\ln[T^{NS}_{c0}/(4\gamma^{}_E
T)]$.

\begin{figure}[ hbtp]
\includegraphics[width=8.5cm]{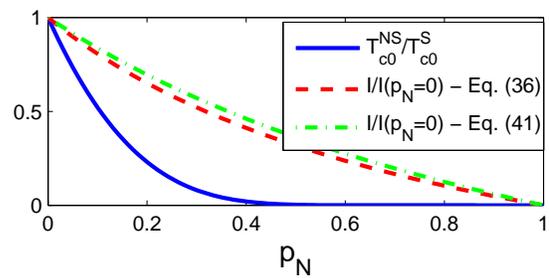}
\caption{(color online) The transition temperature of the bilayer $T_{c0}^{NS}$,
normalized by $T_{c0}^S$ (blue solid line) and the
first harmonic of the current, divided by its value at $p^{}_N=0$, from Eq. (\ref{III}) (red dashed line).
The green dash-dotted line shows the approximation (\ref{approxI}) for the current.
The parameters used for all graphs are $T_{c0}^S=1.27$K (equivalent to $\lambda^{}_S=0.17$), $\omega_D=400$K,
$T_{c0}^N=0$ (i.e. $\lambda^{}_N=0$) and $1/\tau^{}_s=0$.
The current is plotted for $T=4T_{c0}^S$ and $E^{}_c=10T_{c0}^S$.
}\label{fig:IandTc}
\end{figure}

We now present several plots of the persistent current based on
Eq. (\ref{III}). Figure \ref{fig:IandTc} shows the first harmonic
of the current (divided by its value at $p^{}_N=0$) as a function
of $p_N$, for $T^N_{c0}=1/\tau^{}_s=0$. To avoid critical
fluctuations (see below), we restrict ourselves to a relatively
high temperature, $T=4T^S_{c0}$. The same figure also shows the
transition temperature for the bilayer, divided by $T^S_{c0}$.
Clearly, the relative persistent current decreases much more
slowly than the relative transition temperature. This slower
decrease is similar to that found in Refs. \onlinecite{HAMUTAL1}
and \onlinecite{HAMUTAL2}, resulting from the effects of pair
breakers. As an example, for the parameters used in Fig.
\ref{fig:IandTc}, the transition temperature at $p^{}_N=0.7$ is
very small, $T_{c0}^{NS}(p^{}_N=0.7)\simeq 10^{-6}T_{c0}^{S}$
while the first harmonic of the current is given by
$I(p^{}_N=0.7)=0.16I(p_N=0)=0.13E^{}_c$.

\begin{figure}[ hbtp]
\includegraphics[width=8.5cm]{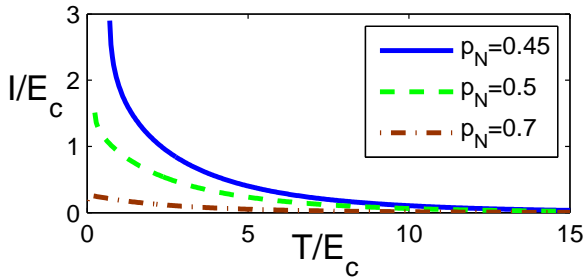}
\caption{(color online) The first harmonic of the current, in
units of the Thouless energy $E_c$, versus $T/E_c$, for
$T_{c0}^S=1.27$K, $E_c=0.015$K, $\omega_D=400$K,
$T_{c0}^N=1/\tau_s=0$ and $\widetilde{{\cal
N}}_{\textrm{eff}}T_{c0}^S=10^5$. The current is plotted for
$p^{}_N=0.45$, $0.5$, and $0.7$ and for $T>T^{NS}_{c0}(p_N)$.
}\label{fig:T_tc0N_is_0_start_TGi}
\end{figure}

At a fixed $p^{}_N$, the persistent current decreases with
increasing temperature. Figure \ref{fig:T_tc0N_is_0_start_TGi}
shows the current (in units of the Thouless energy $E^{}_c$) as a
function of the temperature for a specific choice of the
parameters and for three values of $p^{}_N$. Each of these plots
shows the current only above the transition temperature
$T^{NS}_{c0}$. As anticipated in the Introduction, the persistent
current increases with increasing $E_c$. This can be seen from
Eqs. (\ref{III}) and (\ref{xj}), in which the decay of $I$ is
determined by the ratio $T/E^{}_c$.
 Using the relation $\widetilde{\cal N}_{\rm
eff}T^S_{c0}\equiv g/[E^{}_c/T^S_{c0}]$, where $g$ is the
dimensionless conductance, the parameters used in Fig.
\ref{fig:T_tc0N_is_0_start_TGi} are equivalent to $g=1000$.

\begin{figure}[ hbtp]
\includegraphics[width=8.7cm]{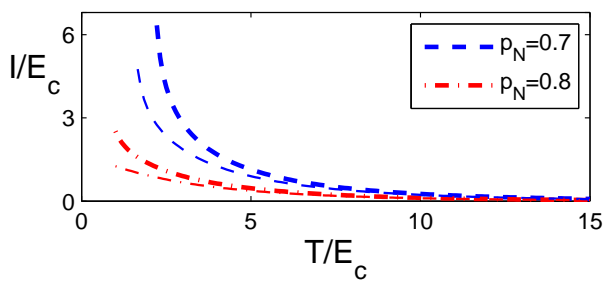}
\caption{(color online) Same as Fig. \ref{fig:T_tc0N_is_0_start_TGi}, but with  $T_{c0}^N=0.001K$ (i.e. $\lambda^{}_N=0.077$). The
current is plotted for  $p^{}_N=0.7$ (blue dashed) and $p^{}_N=0.8$ (red dash-dotted). Thick (thin) lines represent
 $1/\tau_s=0$ ($1/\tau_s=E_c$).
The graphs are plotted for $T>T_{c}^{NS}$. }\label{fig:T_tc0N_finite}
\end{figure}

Finally, we discuss the effects of a positive transition
temperature $T^N_{c0}$ and a finite amount of magnetic impurities
in the normal slab. Figure \ref{fig:T_tc0N_finite} shows the first
harmonic of the persistent current versus the temperature for
$T_{c0}^N=1$mK, which is the estimated minimal value for the pure
transition temperature of copper derived in Ref.
\onlinecite{HAMUTAL2}, with and without magnetic impurities. We
see  that at high temperatures the current is not very sensitive
to the
 pair breaking. As might be expected, the weak superconductivity of the $N$ layer causes an increase in the persistent current.

Interestingly, both Fig. \ref{fig:T_tc0N_is_0_start_TGi} and Fig.
\ref{fig:T_tc0N_finite} exhibit fluctuation-induced persistent
currents which are much larger than the Thouless energy $E^{}_c$,
at temperatures above the superconducting transition temperature
of the $S$ material. This persistent current increases, and its
decay with temperature becomes slower, as $E^{}_c$ increases.

The above plots were based on Eq. (\ref{III}), which sums over
many values of $\ell$ and $j$. For example, Figs.
\ref{fig:T_tc0N_is_0_start_TGi} and \ref{fig:T_tc0N_finite} used
$j,\ell\leq 10^4$. We next describe approximate expressions, which
are valid at high temperatures, when $T\gg T^{NS}_{c0}$. We
restrict this discussion to the case $1/\tau^{}_s=0$. Defining the
small parameter
\begin{align}
w=-1/\ln[T^{NS}_{c0}/(4\gamma^{}_E T)]\ , \end{align} and noting
that $\Psi(w-j)\approx -1/w$, we find $ F^j_{\rm
zero}\approx-j+w$, and therefore
\begin{align}
x^{\ell j}_{\rm zero}\approx x^{\ell j}_{\rm pole}-i m
w\sqrt{\frac{T}{2\pi E^{}_c\bigl[1+2j+|\ell|\bigr]}}\ .
\end{align}
Inserting this approximation into Eq. (\ref{III}), the Poisson
summation form of the current becomes
\begin{widetext}
\begin{align}
I&\approx-4eT\sum_{m=1}^\infty\sin(4\pi m \Phi)
\sum_\ell\sum_{j=0}^\infty\exp[-m\sqrt{2\pi
T[1+2j+|\ell|]/E^{}_c}
\Bigl[\exp\bigl(m w \sqrt{2\pi
T/E^{}_c/[1+2j+|\ell|]}\bigr)-1\Bigr]\nonumber\\
&\approx -4eTw\sum_{m=1}^\infty\sin(4\pi m \Phi)
\sum_\ell\sum_{j=0}^\infty m  \sqrt{2\pi
T/E^{}_c/[1+2j+|\ell|]}\exp[-m\sqrt{2\pi T(1+2j+|\ell|)/E^{}_c}] \
,\label{II}
\end{align}
\end{widetext}
where the last approximation applies only for $m w\sqrt{2\pi
T/E^{}_c}\ll 1$. At intermediate temperatures, when both this
condition and $w\ll 1$ are obeyed,  the current decays as
$I\approx w I^{}_1$, where $I^{}_1$ is independent of
$T^{NS}_{c0}$. Substituting Eq. (\ref{TNS01}) for $T^{NS}_{c0}$ in
the expression for $w$, we finally end up with
\begin{align}
\frac{I}{I^{}_0}\approx \frac{\ln[4\gamma^{}_ET/T^S_{c0}]}{\ln[4\gamma^{}_ET/T^S_{c0}]
+\frac{p^{}_N(1-\lambda^{}_N/\lambda^{}_S)}{p^{}_N\lambda^{}_N+p^{}_S\lambda^{}_S}}\
,\label{approxI}
\end{align}
where $I^{}_0$ is the persistent current (for the same flux) at
$p^{}_N=0$. As seen in Fig. \ref{fig:IandTc}, this approximation is quite good.

Unlike the case of the magnetic impurities, in which the
persistent current remains non-zero even when the transition
temperature vanishes, in the case of the bilayer the persistent
current vanishes when $T^{NS}_{c0}=0$. When $\lambda^{}_N=0$, the
transition temperature approaches zero as $p^{}_N$ increases
towards the quantum critical point, which occurs at $p^{}_N=1$.
When $\lambda^{}_N<0$, this critical point occurs at a threshold
$p^\times_N<1$, and the fluctuation-induced persistent current
vanishes above this threshold. However, as $p^{}_N$ approaches
this critical threshold, the current decreases {\it linearly} with
$p^{}_N$ [as seen from Eq. (\ref{approxI})]. Since the transition
temperature decays exponentially towards that point, we again find
that the persistent current remains significant even when the
transition temperature is negligibly small!

\section{Discussion}
\label{Gin}

The calculations above were carried out within the Gaussian
approximation. This approximation usually breaks down close to the
phase transition, where higher powers of the order parameters must
be taken into account in the expansion of the action ${\cal S}$.
This happens below the so-called Ginzburg temperature,
$T^{}_{Gi}$. Therefore, one should not trust the above results for
temperatures in the range $T^{NS}_c<T<T^{}_{Gi}$. In this range,
we need to supplement Eq. (\ref{ACT2}) by the quartic terms, which
should be derived by continuing the expansion of Eq. (\ref{SSO})
in powers of the $\Delta$'s. 
In principle, this expansion has the
form
\begin{align}
&{\cal S}^{}_4=\frac{1}{2}\beta\widetilde{\cal N}^{}_{\rm
eff}\sum_{\{{\bf
q}^{}_i\}}\sum_{\nu^{}_i}B^{}_{\alpha\beta\gamma\delta}\nonumber\\
&\times\Delta^{}_\alpha({\bf q}^{}_1,\nu^{}_1)\Delta^{}_\beta({\bf
q}^{}_2,\nu^{}_2)\Delta^{}_\gamma({\bf
q}^{}_3,\nu^{}_3)\Delta^{}_\delta({\bf q}^{}_4,\nu^{}_4)\ ,
\label{S4}
\end{align}
where $\alpha,\beta,\gamma,\delta$ take the values $S$ or $N$, and
the sums are restricted by $\sum_i{\bf q}^{}_i=\sum_i\nu^{}_i=0$.

The calculation of the coefficients
$B^{}_{\alpha\beta\gamma\delta}$ goes beyond the scope of the
present paper.\cite{ALTLAND} Usually, these coefficients are
assumed to be independent of the momenta ${\bf q}^{}_i$ and
frequencies $\nu^{}_i$, since such dependencies are less relevant
near the phase transition in the renormalization group sense.
Furthermore, Eq. (\ref{ANS}) shows that the interaction terms in Eq.
(\ref{ACT2}), $|\Delta^{}_S|^2/\lambda^{}_S$ and
$|\Delta^{}_N|^2/\lambda^{}_N$, are equal to their bulk values
multiplied by $p^{}_S$ and $p^{}_N$, respectively. This indicates a renormalization of
$\Delta^{}_S$ and
$\Delta^{}_N$  by the factors $p^{1/2}_S$ and by $p^{1/2}_N$' respectively.
In analogy, we
conjecture that the various coefficients in Eq. (\ref{S4}) are
also given by their bulk values, multiplied by the same renormalization factors.  We next replace these order parameters by
$\Delta^{}_\pm$, from Eq. (\ref{DELPM}). For simplicity,
we restrict the following discussion to the special case
$\lambda^{}_N=0$. This should suffice to demonstrate our
arguments. In this special case one has
$\lambda^{}_{NS}=p^{}_S\lambda^{}_S$,
$a^{}_-=a^{}_S=p^{2}_S[\lambda^{-1}_{NS}-\gamma]$, $u^{}_+=1$ and
$u^{}_-=0$. Also, $a^{}_+=a^{}_N=\infty$, and therefore we can ignore all the fluctuations associated with $\Delta^{}_+$.
Finally, the quartic action becomes
\begin{align}
{\cal S}^{}_4=\frac{1}{2}\beta\widetilde{\cal N}^{}_{\rm
eff}B^{}_-\sum_{\{{\bf q}^{}_i\}}\sum_{\nu^{}_i}\prod_{i=1}^4
\Delta^{}_-({\bf q}_i,\nu_i)\ , \end{align} where again
$\sum_i{\bf q}^{}_i=\sum_i\nu^{}_i=0$ and we set
$B^{}_-=p^2_SB^{}_0$, with $B^{}_0=7\zeta(3)/(8\pi^2T^2)$ having
the bulk value of the quartic term. \cite{VARLAMOV} Keeping only
the first term in Eq. (\ref{ACT3}), we find the usual structure of
the effective Ginzburg-Landau action, except for the
renormalization of the coefficients. As we discuss below, the
dependence of $a^{}_-$ on ${\bf q}$ and on $\nu$ is very different
from the simple form used in standard Ginzburg-Landau theories.

The literature contains several ways to estimate the Ginzburg region.
Since here we calculate the persistent current, we define that
region as the range where the Gaussian calculation presented in
the previous section must be modified by inclusion of the quartic
terms. Expanding the partition function ${\cal Z}$ to leading
order in $B^{}_-$, the free energy becomes
\begin{align}
F=-T\ln{\cal Z}^{}_{\rm fl}+3 \widetilde{\cal N}^{}_{\rm
eff}B^{}_-\Bigl[ \sum_\nu\sum_{\bf q}\langle|\Delta^{}_-({\bf
q},\nu)|^2\rangle\Bigr ]^2\ ,
\end{align}
where $\langle\ldots\rangle$ denotes averaging with the Gaussian
action,
\begin{align}
\langle|\Delta^{}_-({\bf q},\nu)|^2\rangle=1/[\beta\widetilde{\cal
N}^{}_{\rm eff}a^{}_-({\bf q},\nu,T)]\ .
\end{align}
The correction to the persistent current due to the quartic term thus
becomes
\begin{align}
\delta I=\frac{3eB^{}_-T}{\pi\widetilde{\cal N}^{}_{\rm
eff}}\Bigl[\sum_\nu\sum_{\bf
q}\frac{1}{a^{}_-}\Bigr]\Bigl[\sum_\nu\sum_{\bf q}\frac{T\partial
a^{}_-/\partial\Phi}{a^{2}_-}\Bigr]\ .\label{dII}
\end{align}
The above Gaussian results can be used only if this additional contribution is smaller than that calculated
above, Eq. (\ref{II1}).

\begin{figure}[h]
\includegraphics[width=8.5cm]{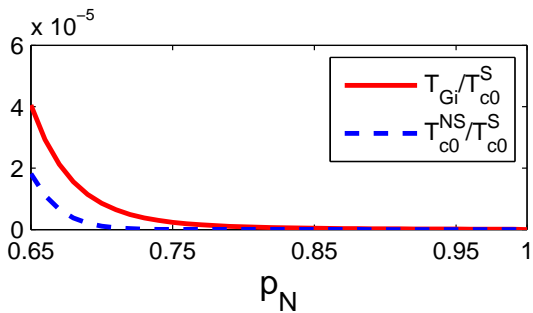}\\
\caption{(color online) $T_{Gi}/T_{c0}^S$ and
$T_{c0}^{NS}/T_{c0}^S$ versus $p_N$, for
$\widetilde{{\cal N}}_{\textrm{eff}}T_{c0}^S=10^5$.  Note the factor of $10^{-5}$ on the
$y$-axis.}\label{fig:TcNSandT_Gi}
\end{figure}

At $|\Phi|=0$, the denominators in the sums in Eqs. (\ref{II1})
and (\ref{dII}) vanish for $n=\ell=0$, at the critical temperature
$T^{NS}_{c0}(\Phi)$ which satisfies Eq. (\ref{shift}). Moving
slightly away from this temperature, i.e. at small
$T-T^{NS}_{c0}(\Phi)$, each of these sums is dominated by its
first term, with $n=\ell=0$. Most of the discussions in the
literature proceed by considering {\it only} these
`zero-dimensional classical' terms.\cite{VARLAMOV} Following this
`tradition', i.e. keeping only these leading terms in all three
sums, and comparing $I$ with $\delta I$, we find that for
$\Phi\approx 0$ the latter can be neglected if
\begin{widetext}
\begin{align}
\frac{1}{\lambda^{}_{NS}}-\frac{1}{\lambda^{}_S}+\Psi\Bigl[\frac{1}{2}+s
p^{}_N\frac{T^S_{c0}}{T}\Bigr]-\Psi\Bigl[\frac{1}{2}
\Bigr]+\ln\Bigl[\frac{T}{T^S_{c0}}\Bigr]>\sqrt{\frac{21\zeta(3)}{4\pi^2T\widetilde{\cal
N}^{}_{\rm eff}p_S^2}}\ ,\label{eqgi}
\end{align}
\end{widetext}
where $s$ was defined after Eq. (\ref{TAG}). The left hand side is
the denominator for $n=\ell=0$, which vanishes at the mean field
transition temperature. Apart from multiplicative factors of order
unity, one obtains a similar `zero-dimensional classical'
condition using other definitions of the Ginzburg
region.\cite{VARLAMOV,FFH} Close to the transition at
$T^{NS}_{c0}$ one usually replaces $T$ by $T^{NS}_{c0}$ in the
denominator of the right hand side. Equation (\ref{eqgi}) then
agrees with the usual Ginzburg criterion in $d$ dimensions, which
would give $(T^{}_{Gi}-T^{}_c)\sim\ln(T^{}_{Gi}/T^{}_c)\sim
T_c^{-2/(4-d)}$, with $d=0$, \cite{VARLAMOV} except for the
additional factor $1/p^{2}_S$. However, this substitution is
problematic when $T^{NS}_{c0}$ is very small, as in our case.
Therefore we prefer to keep $T$ also on the right hand side, and
then solve Eq. (\ref{eqgi}) as an equality.  For $p^{}_N<0.5$ and
for the parameters used above, the resulting $T^{}_{Gi}$ turns out
to be quite close to $T^{NS}_{c0}$, and therefore the
`zero-dimensional classical' Ginzburg region is very narrow.
Therefore, Fig. \ref{fig:TcNSandT_Gi} shows these two temperatures
only for $p^{}_N>0.65$. As seen in this figure, the Ginzburg
temperature does not decay as fast as the transition temperature
at large values of $p^{}_N$. For example, for the value of
$\widetilde{{\cal N}}_{\textrm{eff}}T_{c0}^S$ used in the figure,
we have $T_{Gi}(p_N=0.8)=9*10^{-7}T_{c0}^S$, while
$T_{c0}^{NS}(p_N=0.8)=6*10^{-11}T_{c0}^S$. In any case, the
`classical' Ginzburg region is quite narrow. Had we stopped here
(as done in much of the literature), we would conclude that our
Gaussian results for the persistent current can be used for
practically all temperatures (above $T^{NS}_c$) and relative
widths of the bilayer.

As one moves away from the critical region, each of the sums in
Eqs. (\ref{II1}) and (\ref{dII}) must be supplemented with other
terms, involving both non-zero wave vectors ($n \ne 0$) and
non-zero Matsubara frequencies ($\ell\ne 0$). In fact, we find
that as $n$ increases we need to include more values of $\ell$ to
obtain convergence, and that even the Gaussian approximation, on which most
of our results are based, requires the summation over many classical and
quantum fluctuations. For some reason, most of the literature
ignores the `quantum' fluctuations coming from non-zero $\ell$'s,
and keeps only the `one-dimensional' terms with $n\ne
0$.\cite{VARLAMOV,fink} As discussed above, and in Ref.
\onlinecite{HAMUTAL2}, the persistent current is affected by {\it
both} types of fluctuations. The sum in Eq. (\ref{II1}) always
converges, becoming of order $1/\ln[T/T^S_{c0}]$ for $T^S_{c0}\ll
T\ll E^{}_c$. Similarly, the last sum in Eq. (\ref{dII}) also
converges, becoming of order $1/(\ln[T/T^S_{c0}])^2$. In contrast,
the first sum in Eq. (\ref{dII}) does not converge, and thus it
depends on the cutoffs imposed on the wave vectors $n$ and on the
frequencies $\ell$. This problem arises since our calculation
necessitates the replacement of  the `usual' Green function
$1/(T-T^{}_c+\nu+D{\bf q}^2)$  by $1/[\lambda^{-1}-\gamma({\bf
q},\nu,T)]$, with a logarithmic dependence at large $T$, $\nu$ and $q$ [see e.g. Eq. (\ref{eq.log})].
Since this sum depends on the cutoffs, the resulting Ginzburg
criterion will also depend on these cutoffs.\cite{larkin} In
our case, the dirty diffusive limit imposes the cutoffs
$|\nu|,~D^{}_{\rm eff}{\bf q}^2\ll 1/\tau^{}_+$, where $\tau^{}_+$
is the elastic mean free time.\cite{HAMUTAL2} Replacing the sum
$\sum_{\nu,{\bf q}} (1/a^{}_-)$ by some cutoff dependent constant
still shows that $|\delta I/I|$ decreases with increasing $T$. The
details of this cutoff dependent criterion go beyond the scope of
the present paper.

Our calculation was done for an ensemble of proximity-rings in the
same plane. Qualitatively, we expect similar behavior for two
rings which are deposited on top of each other, which may be
easier to realize experimentally. However, the explicit
calculation for the latter case still needs to be carried out. It
is also interesting to calculate the persistent current when the
bilayer is connected to leads. This also remains for future
calculations.

In conclusion, we have demonstrated that the effect of pair
weakening due to the proximity between a superconducting ring and
a normal (or a weakly superconducting) ring is similar to, but not
identical with, that of pair breaking: the persistent current
decays slowly with the relative width of the normal layer, and
persists even when the superconducting transition temperature
(which decays faster) is very small. Since this relative width can
be controlled, it would be interesting to check our quantitative
predictions experimentally.

\begin{acknowledgments}
We thank A.M. Finkel'stein for a useful discussion. This work was
supported by  the US-Israel Binational Science Foundation (BSF),
by the Israel Science Foundation (ISF) and by its Converging
Technologies Program. OEW and AA acknowledge the support of the
Albert Einstein Minerva Center for Theoretical Physics, Weizmann
Institute of Science, and the hospitality of the Pacific Institute
of Theoretical Physics (PITP) at the University of British
Columbia. JGEH also acknowledges support from NSF grant No.
1106110.
\end{acknowledgments}

\appendix

\section{The partition function}

\label{DETC}

Applying the Hubbard-Stratonovich transformation to Eq.
(\ref{ZZ}), and integrating the fermionic part of the action, the
partition function is cast into the form \cite{ALTLAND}
\begin{align}
{\cal Z}&=\int {\cal D}\{\Delta ({\bf r},\tau ),\Delta^{\ast}_{}({\bf r},\tau )\}e^{-{\cal S}}\ ,\label{ZD}
\end{align}
with the action
\begin{align}
{\cal S}=\int d{\bf r}\int_{0}^{\beta} d\tau\frac{|\Delta ({\bf r},\tau )|^{2}}{V(x)}-\frac{1}{2}{\rm Tr}
\Bigl \{\ln\Bigl (\beta{\cal G}^{-1}\Bigr )\Bigr \}\ .\label{SSO}
\end{align}
Here $\beta $ is the inverse temperature and ${\cal G}^{-1}$ is the inverse
Green function at equal positions and imaginary times,
\begin{align}
{\cal G}^{-1}=\left [\begin{array}{cc}G^{-1}_{p}&i\sigma_{y}\Delta \\ (-i\sigma_y)\Delta^{\ast}&G^{-1}_{h}
\end{array}\right ]\ ,
\end{align}
with
\begin{align}
G^{-1}_{p}&=\left [\begin{array}{cc}-\partial^{}_{\tau}-{\cal H}^{(0)}_{\up\up}&-{\cal H}^{(0)}_{\up\down}\\
-{\cal H}^{(0)}_{\down\up}&-\partial^{}_{\tau}-{\cal
H}^{(0)}_{\down\down}\end{array}\right ]
\end{align}
being the particle inverse Green function, and
\begin{align}
G^{-1}_{h}&=\left [\begin{array}{cc}
-\partial^{}_{\tau}+{\cal H}^{(0)}_{\up\up}&{\cal H}^{(0)}_{\down\up}\\
{\cal H}^{(0)}_{\up\down}&-\partial^{}_{\tau}+{\cal
H}^{(0)}_{\down\down}\end{array}\right ]
\end{align}
being the inverse Green function of the holes. The factor $\beta$
was introduced into the last term in Eq. (\ref{SSO}) to keep the
argument of the log dimensionless (it does not affect any of the
following discussion).

The integration over the bosonic fields in Eq. (\ref{ZD}) is carried out using a stationary-phase
analysis  \cite{ALTLAND} of the action ${\cal S}$. At temperatures above the transition temperature,
this amounts to expanding the second term on the
right-hand side of Eq. (\ref{SSO})  to second order in $\Delta$ (the first-order contribution
 to the expansion being zero)
\begin{align}
&{\rm Tr}\{\ln (\beta {\cal G}^{-1})\}={\rm Tr}\Bigl \{\ln\beta\left [\begin{array}{cc}
G^{-1}_{p}&0 \\ 0&G^{-1}_{h}\end{array}\right ]\Bigr \}\nonumber\\
&+\int \frac{d{\bf r}d{\bf r}'}{\Omega^{2}}\int _{0}^{\beta}
\frac{d\tau d\tau '}{\beta^{2}}{\cal K}({\bf r},{\bf r}',\tau -\tau ' )\Delta ({\bf r}',\tau ')
\Delta^{\ast}_{}({\bf r},\tau )\ ,\label{EXPD}
\end{align}
where $\Omega $ denotes the volume of the system (we added the
factors of volume and $\beta$ to keep ${\cal S}$ dimensionless).
The first term on the right-hand side of Eq. (\ref{EXPD}) will
give the partition function of noninteracting electrons; the
second one represents the contribution of the superconducting
fluctuations to that function. Its calculation requires the
correlation
\begin{align}
{\cal K}({\bf r},{\bf r}',\tau -\tau ')&\equiv - \langle {\rm Tr}\Bigl \{G^{}_{p}({\bf r},{\bf r}',\tau -\tau ')\sigma_{y}\nonumber\\
&\times
G^{}_{h}({\bf r}',{\bf r},\tau '-\tau )\sigma^{}_{y}\Bigr \}\rangle \ ,
\end{align}
where $\langle\ldots\rangle$ indicates averaging over the impurity configurations
(see Ref. \onlinecite{AGD} for details). Upon averaging, the spatial dependence of ${\cal K}$
becomes a function of $x$, $x'$, and $\r -\r '$, where $\r \perp \hat{x}$, see Fig. \ref{first}. Hence,
\begin{align}
&{\rm Tr}\{\ln (\beta {\cal G}^{-1})\}\Big |^{}_{\rm 2^{nd}} \nonumber\\
& = \int \frac{dx dx'}{d^{2}}\sum_{\nu}\sum_{{\bf
q}}\Delta^{}_{x'} ({\bf q},\nu){\cal K}^{}_{xx'}({\bf q},\nu)
\Delta^{\ast}_{x} ({\bf q},\nu)\ ,\label{EXPS}
\end{align}
where
\begin{align}
{\cal K}^{}_{xx'}({\bf q},\nu)&=\sum_{{\bf p}^{}_1,{\bf
p}^{}_2}\sum_{\omega } \langle {\rm Tr}\Bigl \{G^{}_{}(x,x',{\bf
p}^{}_1+{\bf q},{\bf p}^{}_2+{\bf q},\omega+\nu )
\nonumber\\
&\times \sigma_{y} G^{t}_{}(x',x,-{\bf p}^{}_1,-{\bf
p}^{}_2,-\omega)\sigma^{}_{y}\Bigr \}\rangle\ ,\label{KXX}
\end{align}
and both Green functions are the particle one, \cite{HAMUTAL2}
i.e., $G=G_{p}$. [In Eq. (\ref{EXPS}),  $d=d_{N}+d_{S}$ is the
total width of the sandwich.] We use the notations $\omega
\equiv\omega^{}_{n}=\pi T(2n+1)$ for the fermionic Matsubara
frequencies, and $\nu\equiv\nu^{}_{\ell}=\pi T2\ell$ for the
bosonic frequencies. Note that ${\bf q}$, ${\bf p}^{}_1$ and ${\bf
p}^{}_2$ are two-dimensional vectors normal to $x$.

Inserting these results into the expression for the action [see
Eq. (\ref{SSO})], the Gaussian fluctuation-induced partition
function, ${\cal Z}_{\rm fl,2}$, takes the form
\begin{align}
{\cal Z}^{}_{\rm fl,2}=\int {\cal D}\{\Delta^{}_{x} ({\bf
q},\nu),\Delta^{\ast}_{x'} ({\bf q},\nu)\}e^{-{\cal S}^{}_2}\ ,
\end{align}
with
\begin{align}
{\cal S}^{}_2&=\sum_{\bf q}\sum_{\nu}\int dx dx' \Delta^{\ast}_{x}
({\bf q},\nu)\Bigl (\beta\frac{\delta (x-x')}{\widetilde{V}(x)}\nonumber\\
&\hspace{1cm}-\frac{1}{d^2}{\cal K}_{xx'}({\bf q},\nu)
\Bigr )\Delta^{}_{x'}({\bf q},\nu)\ ,
\end{align}
(here $\widetilde{V}$ is the attractive interaction. in units of
energy $\times$ length). In the Cooper limit, the bosonic variable
$\Delta$ takes only two values as a function of $x$,
$\Delta_{N}({\bf q},\nu)$ for $ 0\leq x\leq d_{N}$, and
$\Delta_{S}({\bf q},\nu)$ for $-d_{S}\leq x\leq 0$.  Therefore the
action becomes
\begin{align}
{\cal S}^{}_2=\sum_{\bf q}\sum_{\nu}\Delta^{\dagger}_{}({\bf
q},\nu)\widetilde{\cal S}_{}({\bf q},\nu)\Delta_{}({\bf q},\nu)\
,\label{ACTDD}
\end{align}
where
\begin{align}
&\widetilde{\cal S}_{}({\bf q},\nu)=\beta\left [\begin{array}{cc}d^{}_{N}\widetilde{\cal N}_{N}^{}\lambda_{N}^{-1}&0\\
0&d^{}_{S}\widetilde{\cal N}^{}_{S}\lambda^{-1}_{S}\end{array}\right ]\nonumber\\
&-\frac{1}{d^2}\left [\begin{array}{cc}d^{2}_{N}{\cal K}^{}_{NN}({\bf q},\nu)&
d^{}_{S}d^{}_{N}{\cal K}^{}_{NS}({\bf q},\nu)\\
d^{}_{N}d^{}_{S}{\cal K}^{}_{SN}({\bf q},\nu)&
d^{2}_{S}{\cal K}^{}_{SS}({\bf q},\nu)\end{array}\right ]\ ,\label{ACTD}
\end{align}
$\widetilde{\cal N}_{N(S)}$ denotes the density of states of the
normal (superconducting) layer per unit length,  and
$\Delta^{\dagger}({\bf q},\nu)=\{\Delta^{\ast}_{N}({\bf
q},\nu),\Delta^{\ast}_{S}({\bf q},\nu)\}$.

The functions ${\cal K}_{xx'}({\bf q},\nu)$, Eq. (\ref{KXX}), are
calculated by extending the method employed in Refs.
\onlinecite{DEGENNES} and \onlinecite{ANI} to include the
dependence on $\nu$ and on the two-dimensional wave vector ${\bf
q}$.  The calculation is valid in the dirty limit [in which
$(D/2\pi T)^{1/2}$ is much larger than the mean-free path of the
relevant metal, where $D$ is the diffusion coefficient]. For
simplicity, we omit the vector potential from this calculation;
its effect is incorporated into the result at the end of Sec.
\ref{PSSA}.

We follow the derivation given in Ref. ~\onlinecite{DEGENNES}, and
begin by presenting the response function ${\cal K}$ in the form
\begin{align}
\frac{1}{d^{2}}{\cal K}^{}_{xx'}({\bf q},\nu)&=\sum_{\omega }H^{}_{xx'}({\bf q},\nu,\omega )\ .\label{KH}
\end{align}
Had the normal part of the bilayer filled the entire space, then
 $H^{}_{xx'}=H^{(N)}_{xx'}$, where
\begin{align}
&H^{(N)}_{xx'}({\bf q},\nu,\omega )=\int dq_{x}e^{iq_{x}(x-x')}\nonumber\\
&\times\frac{\widetilde{\cal N}_{N}}{|2\omega| +|\nu|
+2/\tau_{s}+D^{}_{N}({\bf q}^{2}+q^{2}_{x})}\ .\label{HN}
\end{align}
We have allowed for
scattering off magnetic impurities in this metal, whose effect is presented by the spin-flip rate $1/\tau_{s}$.
(The effect of scattering off nonmagnetic impurities is contained in the diffusion coefficient.)
As seen from Eq. (\ref{HN}), the function $H^{(N)}_{xx'}$ obeys a diffusion equation
\begin{align}
&\Bigl (|2\omega |+|\nu| +2/\tau_{s}+D^{}_{N}{\bf q}^{2}\nonumber\\
&-D^{}_{N}\frac{\partial^{2}}{\partial x'^{2}}\Bigr
)H^{(N)}_{xx'}({\bf q},\nu,\omega ) =2\pi\widetilde{\cal
N}^{}_{N}\delta (x-x')\ .
\end{align}
Quite similarly, when the $S$ metal fills the entire space one finds
\begin{align}
&\Bigl (|2\omega | +|\nu| +D^{}_{S}{\bf q}^{2}\nonumber\\
&-D^{}_{S}\frac{\partial^{2}}{\partial x'^{2}}\Bigr
)H^{(S)}_{xx'}({\bf q},\nu,\omega ) =2\pi\widetilde{\cal
N}^{}_{S}\delta (x-x')\ .
\end{align}
Here it was assumed that the $S$ metal is not doped with magnetic impurities.
It follows that in order to find $H_{xx'}$ of the double layer, one has to solve the set of equations
\begin{widetext}
\begin{align}
\Bigl (|2\omega |+|\nu| +2/\tau_{s}+D^{}_{N}{\bf q}^{2}
-D^{}_{N}\frac{\partial^{2}}{\partial x'^{2}}\Bigr )H^{}_{xx'}({\bf q},\nu,\omega )
=2\pi\widetilde{\cal N}^{}_{N}\delta (x-x')\ ,\ \ \ x'>0\ ,\nonumber\\
\Bigl (|2\omega | +|\nu| +D^{}_{S}{\bf q}^{2}
-D^{}_{S}\frac{\partial^{2}}{\partial x'^{2}}\Bigr
)H^{}_{xx'}({\bf q},\nu,\omega ) =2\pi\widetilde{\cal
N}^{}_{S}\delta (x-x')\ ,\ \ x'<0\ ,
\end{align}
\end{widetext}
with the appropriate boundary conditions.
Such a scheme has been undertaken in Refs. \onlinecite{DEGENNES}
and \onlinecite{ANI}, leading to the result
\begin{align}
&H^{}_{NN}({\bf q},\nu,\omega)=\frac{\pi\widetilde{\cal
N}_{N}^{2}}{d^{}_{N}\widetilde{\cal N}^{}_{N}
+d^{}_{S}\widetilde{\cal N}^{}_{S}}\bar{\gamma}^{}_{}({\bf q},\nu,\omega)\ ,\nonumber\\
&H^{}_{NS}({\bf q},\nu,\omega)=H^{}_{SN}({\bf q},\nu,\omega)
=\frac{\pi\widetilde{\cal N}_{N}^{}\widetilde{\cal
N}^{}_{S}}{d^{}_{N}\widetilde{\cal N}^{}_{N}
+d^{}_{S}\widetilde{\cal N}^{}_{S}}\bar{\gamma}^{}_{}({\bf q}, \nu, \omega)\ ,\nonumber\\
&H^{}_{SS}({\bf q},\nu,\omega)=\frac{\pi\widetilde{\cal
N}_{S}^{2}}{d^{}_{N}\widetilde{\cal N}^{}_{N}
+d^{}_{S}\widetilde{\cal N}^{}_{S}}\bar{\gamma}^{}_{}({\bf q},
\nu, \omega)\ ,\label{H01}
\end{align}
where $\bar{\gamma}^{}_{ }({\bf q},\nu,\omega)$ was defined in Eq.
(\ref{GAM}).

Inserting Eqs. (\ref{KH}) and  (\ref{H01}) into  Eq. (\ref{ACTD})
brings the action ${\cal S}$, Eq.  (\ref{ACTDD}), into the form
(\ref{ACT2}).


\begin{thebibliography}{999}

\bibitem{BUTTIKER}

M. B\"{u}ttiker, Y. Imry, and R. Landauer, Phys. Lett.  {\bf 96A}, 365 (1983).

\bibitem{BOOK}
Y. Imry, {\it Introduction to Mesoscopic Physics},  2nd ed. (Oxford University Press, Oxford, 2002).



\bibitem{AEEPL}
V. Ambegaokar and U. Eckern, Europhys. Lett. {\bf 13}, 733
(1990).

\bibitem{AEPRL}
V. Ambegaokar and U. Eckern, Phys. Rev. Lett. {\bf 65}, 381
(1990).

\bibitem{LDDB}
L. P. Levy, G. Dolan, J. Dunsmuir, and H. Bouchiat, Phys. Rev.
Lett. {\bf 64}, 2074 (1990).

\bibitem{reulet} B. Reulet, M. Ramin, H. Bouchiat, and D. Mailly, Phys. Rev. Lett. {\bf 75}, 124 (1995).

\bibitem{DBRBM}
R. Deblock, R. Bel, B. Reulet, H. Bouchiat, and D. Mailly, Phys.
Rev. Lett. {\bf 89}, 206803 (2002).


\bibitem{webb91}
V. Chandrasekhar, R. A. Webb, M. J. Brady, M. B. Ketchen, W. J.
Gallagher, and A. Kleinsasser, Phys. Rev. Lett. {\bf 67}, 3578
(1991).

\bibitem{10} D. Mailley, C. Chapelier, and A. Benoit, Phys. Rev. Lett. {\bf 70}, 2020 (1993).

\bibitem{11} W. Rabaud, L. Saminadayar, D. Mailly, K. Hasselbach, A. Benoit, and B. Etienne, Phys. Rev. Lett. {\bf 86}, 3124 (2001).

\bibitem{Moler}
H. Bluhm, N. C. Koshnick, J. A. Bert, M. E. Huber, and K. A. Moler,
Phys. Rev. Lett. {\bf 102}, 136802 (2009).


\bibitem{JMKW}
E. M. Q. Jariwala, P. Mohanty, M. B. Ketchen, and R. A. Webb,
Phys. Rev. Lett. {\bf 86}, 1594 (2001).


\bibitem{JGEH}

A. C. Bleszynski-Jayich, W. E. Shanks, B. Peaudecerf, E. Ginossar,
F. von Oppen, L. Glazman, and J. G. E. Harris, Science {\bf 326},
272  (2009).

  \bibitem{EG}
E. Ginossar, L. I. Glazman, T. Ojanen, F. von Oppen, W. E.
Shanks, A. C. Bleszynski-Jayich, and J. G. E. Harris, Phys. Rev.
B {\bf 81}, 155448 (2010).



\bibitem{HAMUTAL3}


See also e.g. H. Bary-Soroker, O. Entin-Wohlman, and Y. Imry, Phys.
Rev. B {\bf 82}, 144202 (2010).

\bibitem{GD0}
 C. Vallette,
Solid State Comm. {\bf 9}, 895 (1971).

\bibitem{GD2}
G. Deutscher, S. Y. Hsieh, P. Lindenfeld, and S. Wolf, Phys. Rev. B {\bf 8}, 5055 (1973).

\bibitem{GD1}
G. Deutscher, Solid State Comm. {\bf 9}, 891 (1971).

\bibitem{HELENE}


H. Bouchiat, Physics {\bf 1}, 7 (2008).






\bibitem{HAMUTAL1}


H. Bary-Soroker, O. Entin-Wohlman, and Y. Imry, Phys. Rev. Lett.
{\bf 101}, 057001 (2008).

\bibitem{HAMUTAL2}

H. Bary-Soroker, O. Entin-Wohlman, and Y. Imry, Phys. Rev. B {\bf
80}, 024509 (2009).

\bibitem{BIRGE}
F. Pierre, A. B. Gougam, A. Anthore, H. Pothier, D. Esteve, and
N. O. Birge, Phys. Rev. B {\bf 68}, 085413 (2003).

\bibitem{oreg} G. Schwiete and Y. Oreg, Phys. Rev. Lett. {\bf
103}, 037001 (2009); Phys. Rev. B {\bf 82}, 214514 (2010).

\bibitem{DEGENNES}
P. G. de Gennes, Rev. Mod. Phys. {\bf 36}, 225 (1964).

\bibitem{ANI}

O. Entin-Wohlman, Phys. Rev. B {\bf 12}, 4860 (1975).



\bibitem{FULDE}


P. Fulde, in {\it Bardeen Cooper and Schrieffer: 50 years}, eds. L. N.
Cooper and D. Feldman (World Scientific, 2010).




\bibitem{COM2}



The coherence length, in the dirty limit, is given by
$\xi_{N,S}\propto \sqrt{D_{N,S}/(2\pi T)}$; the dirty-limit
condition is $\ell_{N,S}<<\xi_{N,S}$,  where $\ell_{N,S}$ is the
elastic mean-free path.



\bibitem{MAC}

W. L. McMillan, Phys. Rev. {\bf 175}, 537 (1968).





\bibitem{AG}
A. A. Abrikosov and L. P. Gorkov, Soviet Physics JETP {\bf 12},
1243 (1961).






\bibitem{AGD}
A. A. Abrikosov, L. P. Gorkov, and I. E. Dzyaloshinskii, {\it
Methods of Quantum Field Theory in Statistical Physics}
(Prentice-Hall, Englewood Cliffs, NJ, 1963).


\bibitem{ALTLAND}

A. Altland and B. Simons, {\it Condensed Matter Field Theory}
(Cambridge University Press, Cambridge, 2006).


\bibitem{carolimaki}

C. Caroli and K. Maki, Phys. Rev. {\bf 159}, 306 (1967).


\bibitem{VARLAMOV}


A. I. Larkin and A. Varlamov, {\it Theory of Fluctuations in
Superconductors} (Oxford University Press, 2009).

\bibitem{FFH}
D. S. Fisher, M. P. A. Fisher and D. Huse, Phys. Rev. B {\bf 43},
130 (1991).

\bibitem{fink} An exception was treated by K. Michaeli and A. M.
Finkel'stein, Phys. Rev. B {\bf 80}, 214516 (2009).


\bibitem{larkin} See e.g. V. M. Galitsky and A. I. Larkin, Phys. Rev. B {\bf 63}, 174506 (2001); V. M. Galitski,
Phys. Rev. Lett. {\bf 100}, 127001 (2008).

\end{thebibliography}
\end{document}